\documentclass[sigconf,nonacm]{acmart}
\newtoggle{arxiv_submission}
\toggletrue{arxiv_submission}

\makeatletter
\iftoggle{arxiv_submission}{%
  \renewenvironment{figure}[1][tbp]{\@dblfloat{figure}[#1]}{\end@dblfloat}
  \setlength{\emergencystretch}{3em}
}{}
\makeatother

\newcommand{\sourcecodefootnote}{%
  \iftoggle{arxiv_submission}{%
    Source code: \url{https://github.com/a554b554/Reactant}.%
  }{%
    For review, the source code is available in the supplementary materials.%
  }%
}

\AtBeginDocument{%
  }

\setcopyright{none}
\acmDOI{}
\acmISBN{}

\usepackage{listings}
\usepackage[most]{tcolorbox}
\usepackage{tabularx}
\definecolor{caseTeal}{HTML}{087E83}
\definecolor{casePurple}{HTML}{6C4AA1}
\definecolor{caseOrange}{HTML}{9A4F14}
\definecolor{caseBlue}{HTML}{245CB2}
\definecolor{caseOlive}{HTML}{526918}
\definecolor{caseRed}{HTML}{A82E3D}
\definecolor{caseBase}{HTML}{374151}
\definecolor{caseManual}{HTML}{64748B}
\tcbset{casefinal/.style={colback=black!2,colframe=black!15,boxrule=.5pt,arc=2pt,
  left=8pt,right=8pt,top=6pt,bottom=6pt,coltitle=black,colbacktitle=black!6,
  fonttitle=\sffamily\bfseries\small,before skip=2pt,after skip=5pt}}

\definecolor{codebg}{gray}{0.93}
\definecolor{tagcolor}{HTML}{0969DA}
\definecolor{protectcolor}{HTML}{E42318}
\definecolor{fieldcolor}{HTML}{D97706}
\definecolor{outputcolor}{HTML}{0FA76E}
\definecolor{refcolor}{HTML}{7C3AED}

\lstdefinestyle{basecode}{
  basicstyle=\ttfamily\small,
  breaklines=false,
  columns=fullflexible,
  keepspaces=true,
}

\lstdefinestyle{reactant}{
  style=basecode,
  moredelim=*[s][\color{tagcolor}\bfseries]{<@}{>},
  moredelim=*[s][\color{outputcolor}\bfseries\itshape]{<@output}{>},
  moredelim=**[s][\color{protectcolor}\bfseries]{<<}{>>},
  moredelim=**[s][\color{fieldcolor}]{((}{))},
  moredelim=**[s][\color{refcolor}]{``}{``},
}

\newtcblisting{codeblock}[1][style=reactant]{
  colback=codebg,
  colframe=codebg,
  listing only,
  listing options={#1},
  boxrule=0pt,
  arc=0pt,
  left=6pt, right=6pt, top=2pt, bottom=2pt,
}

\begin{document}

\title[Who Asked for This?]{Who Asked for This? Inline Annotations as Authoring Transactions for Provenance in Agentic Authoring}

\author{Chang Xiao}
\affiliation{%
  \institution{Boston University}
  \city{Boston}
  \country{USA}}
\email{xchang@bu.edu}

\begin{abstract}
Writing with AI agents turns a paragraph into the outcome of many requests, yet the finished document rarely explains which request produced which change. We introduce Reactant, an interaction paradigm in which authors place typed inline annotations in their original documents. A verified transaction protocol records each request, skill identity, and the correspondences that identify additions, deletions, and transformations. The kernel validates the witness against the recorded states to establish word-level longitudinal lineage. We demonstrate Reactant through this paper's revision history, and report four months of three colleagues' self-directed use. Their uses include conversational inquiry into history and deriving reusable skills from recurring requests, illustrating how the transaction record serves as an extensible substrate for agentic authoring.

\end{abstract}


\maketitle

\section{Introduction}

\begin{figure}[!t]
    \centering
    \includegraphics[width=\linewidth]{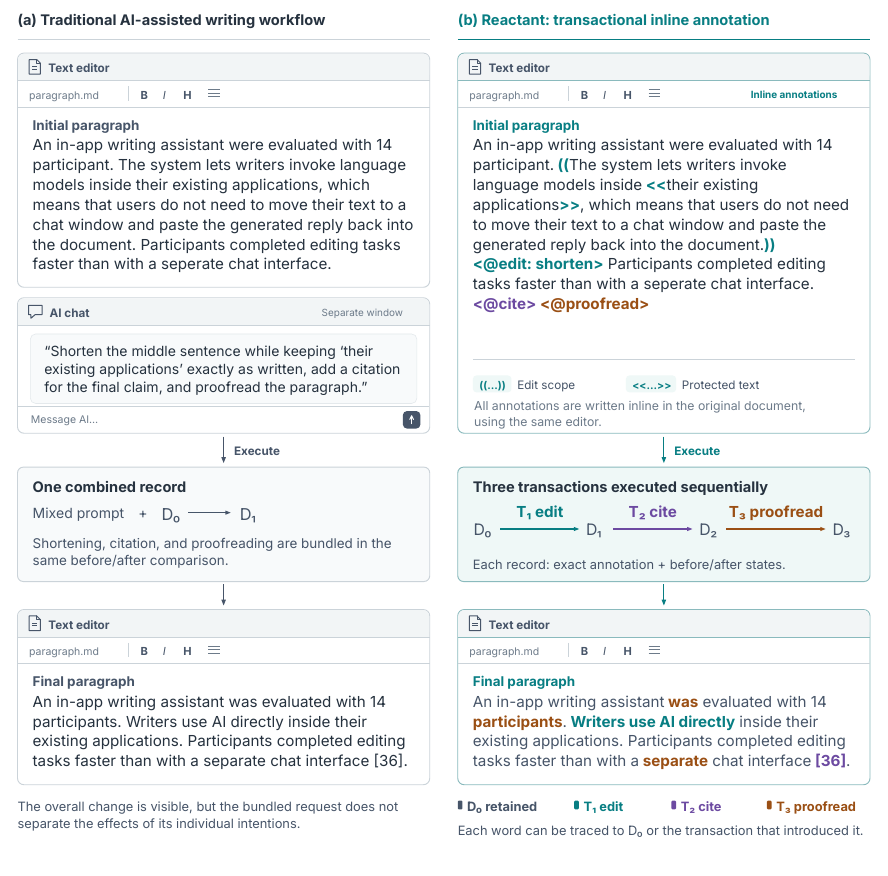}
    \caption{Compared to the traditional AI-assisted writing workflow, Reactant processes inline annotations in order, recording each request and its exact before/after states as a transaction with a validated structured edit witness. The scope modifiers \texttt{((...))} limit the edit and \texttt{\textless\kern0pt\textless...\textgreater\kern0pt\textgreater} protect exact text. Final-text colors identify retained words and the operations that introduced new wording. The comparison concerns the illustrated requests; conversational tools may retain additional history.}
    \Description{Two columns show the same three-sentence paragraph about a writing assistant evaluated with 14 participants. The initial paragraph contains three errors: ``assistant were,'' ``14 participant,'' and ``seperate.'' On the left, a separate chat combines shortening the middle sentence, preserving a phrase, adding a citation, and proofreading. On the right, all three inline annotations are present together, with proofread at the end of the paragraph. Double parentheses delimit the sentence to shorten, and double angle brackets protect the phrase ``their existing applications.'' A single execution processes edit, cite, and proofread as three recorded transactions. Both outputs are identical. In Reactant's final paragraph, teal identifies the new words ``Writers use AI directly'' from the first transaction, purple identifies the citation added by the second transaction, orange identifies ``was,'' ``participants,'' and ``separate'' corrected by the final proofreading transaction, and gray identifies retained words.}
    \label{fig:request-granularity}
\end{figure}

Writing is among the most common tasks people delegate to AI: an OpenAI study of ChatGPT usage found that writing is the most common work-related task, accounting for about 40\% of work messages~\cite{chatterji_how_2025}. With recent advances in agentic AI, AI writing assistance is becoming agentic as well. Beyond chat assistants that return text for authors to paste back, writing tools now offer agents that act on documents directly: specialized agents in Grammarly~\cite{grammarly_agents_2025}; Agent Mode in Microsoft Word~\cite{microsoft_vibe_2025}; shared agents in collaborative editors~\cite{lehmann_collaborative_2026}; and general-purpose agents such as Claude Code, Codex, and Cursor that read and edit local documents with data and code as context~\cite{anthropic_claude_2025,openai_codex_2025,anysphere_cursor_2024}. 

While this could increase efficiency in certain aspects of writing, such as validating a claim against experimental data or locating relevant citations, it also turns the paragraph into the outcome of many small, heterogeneous transformations: the author drafts it, asks AI to proofread it, and thinks about rewording a sentence to make it clearer, and asks AI for a citation, and finally shortens it with AI. Each request has an intent, a targeted place, and snapshots of the text before and after. However, what we see is only the final state. When a coauthor later meets a sentence they do not recognize and asks \emph{who asked for this}, nothing in the document can answer.

This loss matters because people later misremember whether ideas and text came from themselves or AI, with correct attribution declining most sharply in mixed human--AI workflows~\cite{zindulka_ai_2026}. Beyond memory, writers seek control over the aspects of composition they consider their primary contribution~\cite{reza_co-writing_2025}. Prior work shows that making AI contributions visible helps writers reflect on their use of AI and retain a sense of control and ownership~\cite{hoque_hallmark_2024}. Edits also carry social meaning in collaborative writing~\cite{birnholtz_tracking_2012}, so preserving the intent behind an agent's changes could help coauthors interpret them. Some academic journals and conferences also require authors to disclose substantive uses of generative AI~\cite{elsevier_generative_2026,recsys_ai_2026}, while emerging frameworks call for detailed accounts of how and where AI intervened and how its output was reviewed~\cite{xexeo_faceted_2026}. These needs motivate provenance linking requests to affected passages and resulting changes across revisions. A binary ``human'' or ``AI'' label cannot explain how proofreading, revising, and rewriting successively shaped the same passage.

Academic systems and products, including HaLLMark~\cite{hoque_hallmark_2024}, DraftMarks~\cite{siddiqui_draftmarks_2026}, Grammarly Authorship~\cite{grammarly_authorship_2024}, Claude Code checkpoints~\cite{anthropic_checkpoints_2026}, and aider~\cite{aider_watch_2025}, support provenance in AI-assisted writing. However, two limitations motivate our approach. First, many writing-provenance tools tie their features to dedicated authoring environments or editors, requiring authors who use other editors to leave their preferred environment and adopt a different interface. Second, tracing individual contributions often requires inspecting execution histories or file diffs after the fact. This becomes difficult when a single prompt combines multiple intentions whose effects are intertwined in the resulting revision, and the final changes may be drastic, making it hard to establish word-level lineage by simply comparing versions. We therefore need a mechanism that captures authoring intentions as distinct requests at the input stage, and preserves those boundaries through execution, reducing reliance on post hoc reconstruction.

Our observation comes from a familiar practice in human collaborative writing: authors place requests for change next to the text they concern. In \LaTeX{}, packages such as \texttt{todonotes} support TODO notes within the manuscript~\cite{midtiby_todonotes}. PDF readers and collaborative editors such as Overleaf and Google Docs similarly let authors select text and place comments for collaborators to address. Such a comment expresses an intended change and indicates the passage it concerns. We view revising that passage in response to the comment as an \textit{authoring transaction}: the comment supplies the request and its scope, and the revision has a document state before and after it. Preserving these elements together would let collaborators revisit exactly what was requested, inspect the resulting change even after the comment is resolved, and understand how the document has evolved in response to these comments.

This observation suggests extending the same practice to AI-assisted writing. We introduce \textit{Reactant}, an interaction paradigm in which authors write inline annotations directly in the document they want to revise, using their existing editor and source format, such as \LaTeX{} or Markdown. Each annotation names an operation and can include a natural-language instruction, with its placement and optional range markers indicating the intended scope. A compatible general-purpose agent (e.g., Claude Code, Codex) reads the annotation in context and carries out the requested revision. Each annotation thus supplies a contextual unit of delegation.

Reactant implements these delegations through a \emph{verified transaction protocol}. An authoring transaction retains the exact request, skill identity, and before/after document states. After execution, the agent submits a \emph{structured edit witness}: a payload containing \emph{structured edit correspondences} that identify added and deleted spans and link transformed wording to its predecessor. The kernel validates the witness against the recorded states before retaining it and using its correspondences to establish \emph{longitudinal lineage}. These links connect successive revisions even when wording changes, allowing authors to inspect both earlier formulations and the requests that transformed them.

Figure~\ref{fig:request-granularity} shows two workflows producing the same revision from the same initial text. Reactant records three inline operations as separate transactions, preserving each instruction and intermediate revision within a single execution.

Beyond provenance and lineage, this design offers three practical advantages. First, authors draft text, place instructions, and receive revisions in the same document, avoiding manual transfer between the draft and a separate chat interface. This design responds to findings that writers perceive leaving the primary editing context during AI-assisted revision as effortful and disruptive~\cite{reza_abscribe_2024}. Second, the protocol is independent of any particular editor: authors can continue using their preferred editor, provided that a compatible AI agent can access the document. This aligns with prior findings that providing LLM assistance within existing applications can support faster editing and higher perceived usability than a separate chat interface~\cite{teufelberger_llmforx_2024}. Third, the annotation grammar is extensible. Reactant supplies built-in operations for proofreading, filling placeholders, and adding citations, while authors can customize and define new ones easily through natural language in our protocol. These extensions use the same annotation syntax and transaction mechanism, so new operations also contribute to the document's provenance history.

This paper makes three contributions:
\begin{itemize}
    \item \textbf{An interaction paradigm for agentic writing.} Authors express requests through transactional inline annotations within their existing documents, making each annotation a contextual unit of delegation.
    \item \textbf{A verified transaction protocol for longitudinal lineage.} The Reactant kernel records requests and exact before/after states, accepts structured edit witnesses from executing agents, and validates these witnesses against the recorded states to establish lineage across revisions.
    \item \textbf{An extensible substrate demonstrated through real-world use.} Four months of use and user-developed extensions demonstrate how transaction histories support functionality beyond revision inspection, including conversational inquiry and the extraction of reusable skills from recurring requests.
\end{itemize}

Additionally, our system is fully open-sourced\footnote{\sourcecodefootnote} and readily available to be installed directly as a plug-in on major agent platforms, such as Claude Code, Codex, or OpenCode.

\section{Related Work}\label{sec:related_work}

\paragraph{AI-assisted Writing Tools}
HCI researchers have long studied how writing assistants can help people write more effectively, developing systems that support diverse tasks and forms of interaction across the writing process~\cite{lee_design_2024,reza_co-writing_2025}.
Soylent~\cite{bernstein_soylent_2010} delegates text editing tasks, such as shortening and proofreading, to crowd workers; ABScribe~\cite{reza_abscribe_2024} turns AI prompts into reusable modifiers for exploring writing variations; and InkSync~\cite{laban_beyond_2024} presents AI-suggested executable insertion, replacement, and deletion suggestions within a document.
DirectGPT~\cite{masson_directgpt_2024} and Script\&Shift~\cite{siddiqui_scriptshift_2025} further connect AI operations to direct manipulation and layered interfaces.
Commercial tools such as Gemini in Docs~\cite{google_gemini_2024} and Notion AI~\cite{notion_what_2026} likewise offer in-app AI-assisted writing features for various demands.
However, these capabilities depend on particular editors or platforms.
Writers who already use other editing environments may need to switch tools or transfer text between interfaces, creating a barrier to adoption within established workflows.

A practical alternative is to ask a general-purpose agent such as Claude Code~\cite{anthropic_claude_2025} or Cursor~\cite{anysphere_cursor_2024} to operate on the document directly.
Because a general AI agent can browse the local project folder, it can obtain the context easily and process more complex requests.

Beyond completing requests, AI writing assistance needs to preserve authors' control over how AI contributes to their work.
Reza et al.~\cite{reza_co-writing_2025} found that authors seek different levels of AI involvement across the writing process, prioritizing control over the parts they consider their primary contribution.
Zindulka et al.~\cite{zindulka_ai_2026} showed that people misremembered whether ideas and text had been created with AI after one week, with the steepest decline in correct attribution in mixed human--AI workflows.
These findings motivate interfaces that let authors specify what to delegate and revisit how AI shaped their text.
Provenance can support this reflection by preserving an account of authors' requests and the resulting changes, reducing the need to reconstruct AI involvement from memory alone.

\paragraph{Provenance and Revision Histories in AI-Assisted Writing}
Writing-process research has long captured activity beyond the finished document: Inputlog~\cite{leijten_inputlog_2013} records keystrokes and revision activity for analysis.
Contemporary tools extend such capture to human--AI writing.
Grammarly Authorship~\cite{grammarly_authorship_2024} distinguishes typed, pasted, and AI-assisted text within supported applications and provides a replay, while Humanly~\cite{zhu_humanly_2026} records activity and in-platform AI assistance in configurable writing environments and packages sessions into writing certificates.
DraftMarks~\cite{siddiqui_draftmarks_2026} captures versioned editor states through an editor-state listener across several writing configurations and exposes revision traces and phrase-level provenance to readers.
History Flow~\cite{viegas_history_flow_2004} visualizes text survival and contributor relationships across revisions, while DocuViz~\cite{wang_docuviz_2015} uses Google Docs revision histories to reveal collaborative writing patterns.
CoAuthorViz~\cite{shibani_coauthorviz_2023} examines the use and modification of GPT-3 suggestions in CoAuthor logs.
HaLLMark~\cite{hoque_hallmark_2024} links prompts to adopted text and distinguishes editorial from generative assistance; InkSync~\cite{laban_beyond_2024} retains a trace of generated content for later factual verification.

Git and agent checkpoints provide another basis for retaining requests and revisions.
Claude Code checkpoints~\cite{anthropic_checkpoints_2026} associate file snapshots with user prompts, while aider can collect several inline instructions in one trigger and retain resulting changes through Git~\cite{aider_watch_2025}.
A conversational turn or trigger can therefore group multiple writing decisions into one revision.
Retaining that history does not by itself guarantee a distinct record for each request, making the boundary used to group changes consequential for later attribution.
Creative version-control research similarly highlights practitioners' interest in histories that preserve meaning and intent alongside states~\cite{sterman_towards_2022}.

These works motivate preserving both request boundaries and explicit transformation links. Reactant executes and records each typed inline annotation separately, retaining the exact request, skill identity, and before/after states alongside an agent-submitted structured edit witness. The kernel validates the witness's edit correspondences against these states before establishing longitudinal lineage. This verified transaction protocol supplies a shared substrate that compatible agents and custom skills can extend and reuse.

\section{System Implementation}\label{sec:system}

Reactant uses a general-purpose AI agent (e.g., Claude Code, Codex) to interpret and execute writing requests through inline operation-specific skills. A Python kernel implements the verified transaction protocol: it records each request and the exact before/after states, validates the agent-submitted structured edit witness against those states, and retains its edit correspondences to establish longitudinal lineage at word level. New skills use this same protocol to submit witnesses and record their operations, so authors can extend Reactant without implementing separate provenance storage or lineage logic. The record preserves the declared request and observed change, without establishing that the agent understood the request or produced correct text.

\subsection{Writing Requests in the Document}

As shown in Figure~\ref{fig:request-granularity}, an author places an annotation such as \texttt{<@edit: shorten>} beside the text to revise, then invokes \texttt{/reactant:execute \$filename} in a generic AI tool like Claude Code with our Reactant plugin installed. We use \emph{agentic writing} for this practice of delegating document transformations to an agent that can read and modify the document itself. Reactant currently operates on local UTF-8 text, including \LaTeX{}, Markdown, and plain text. Reactant does not require the editor to expose its keystrokes, selections, or internal document model.

An annotation has the form \texttt{<@[skill name]: [prompt]>}, where the bracketed terms are placeholders for the skill to invoke and its natural-language instruction. The colon and prompt may be omitted, yielding \texttt{<@[skill name]>}. Its placement supplies local context. For the editing skill, the default block is the maximal run of nonblank lines containing the annotation, while an explicit instruction can identify a different passage. Optional field markers, \texttt{((...))}, restrict the editable region, and protection markers, \texttt{\textless\kern0pt\textless...\textgreater\kern0pt\textgreater}, require exact preservation. Double-backtick references identify additional files to read or destinations to write (e.g., \verb|``results.csv``| for input data ).

\begin{table}[tb]
\caption{Reactant's eight built-in skills. Each is activated by a router-selected annotation and uses the same transaction boundary when history is enabled. The three entry skills are separate from this annotation namespace.}
\label{tab:skills}
\small
\begin{tabularx}{\linewidth}{@{}lX@{}}
\toprule
Skill & Behavior \\
\midrule
\texttt{edit} & Generic request. Revise the surrounding block according to the prompt. \\
\texttt{proofread} & Correct grammar, spelling, and punctuation while preserving meaning and sentence structure. \\
\texttt{ph} & Placeholder tag. Fill it with content inferred from its instruction and context. \\
\texttt{cite} & Find supporting sources and insert citations; in \LaTeX{}, update the bibliography as needed. \\
\texttt{plan} & Answer the author's inline chat turn by appending an AI response immediately after it, without changing the original content. The AI's output has an output tag. \\
\texttt{resolve} & Apply a completed chat chain's agreed revisions to the associated passage and remove the discussion chain. \\
\texttt{figure} & Generate an image using an available host capability and insert its reference. \\
\texttt{plot} & Generate a visualization from supplied data and retain the plotting script and output. \\
\bottomrule
\end{tabularx}
\end{table}

\subsection{A Holistic Authoring Example}\label{sec:holistic-example}

\begin{figure}[t]
\centering
\includegraphics[width=\linewidth]{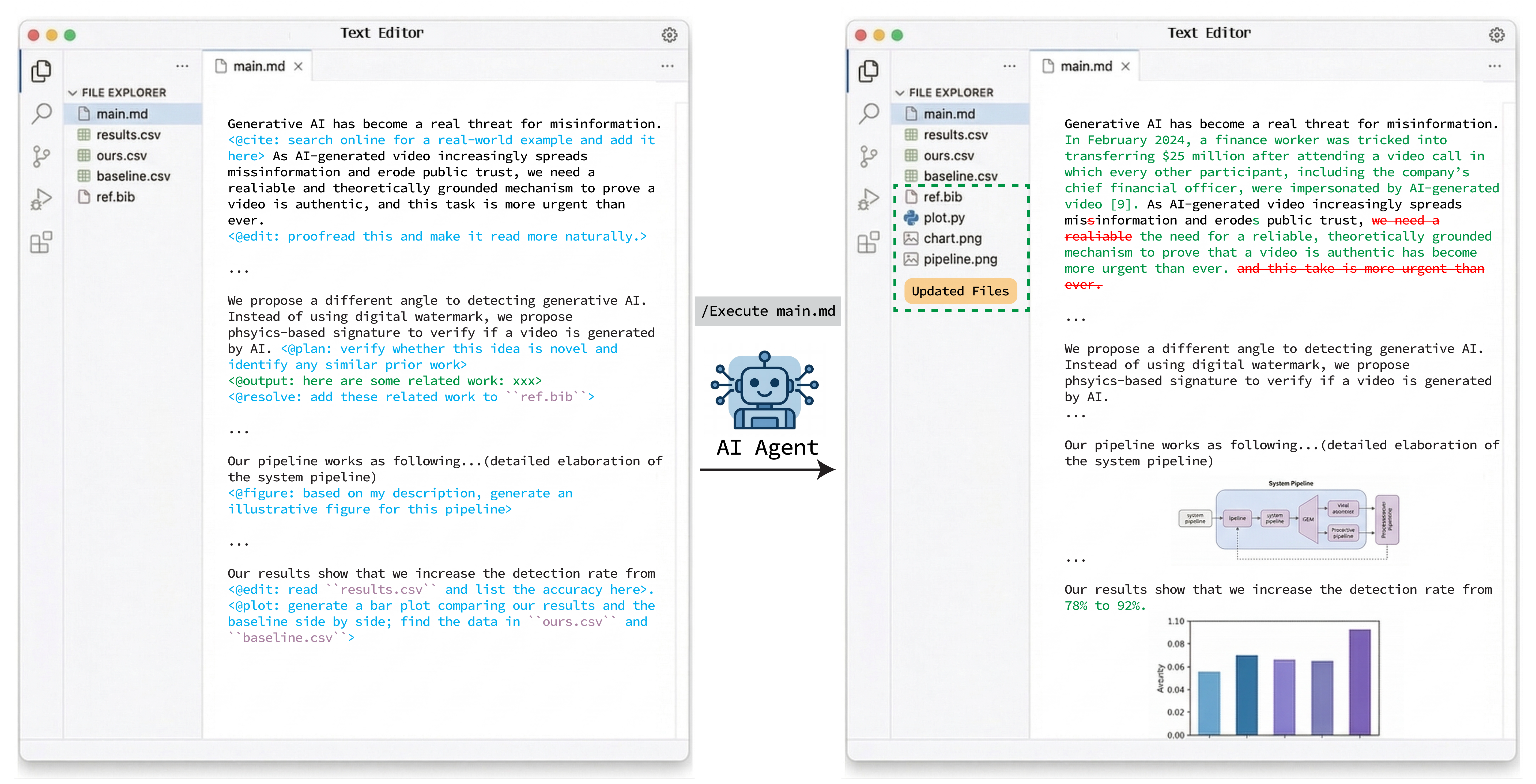}
\caption{A holistic report-writing example. Inline requests appear beside the relevant passages (left). Execution updates the report and supporting files (right). Green and red text illustrate additions and deletions.}
\Description{Two editor views show a report before and after inline requests are executed. The opening paragraph requests a supporting example and a prose revision. A discussion beside the proposed method is followed by a resolution request. Later requests generate a pipeline illustration, fill in results from a CSV file, and plot a comparison from two data files. The updated view highlights text changes and lists a bibliography, plotting script, chart, and pipeline image.}
\label{fig:teaser}
\end{figure}

Figure~\ref{fig:teaser} illustrates Reactant in a Markdown report on detecting AI-generated videos, drawn from one author's work. Inline requests combine citation, prose revision, discussion, and artifact generation, updating the report alongside its bibliography, figures, and plotting script. The figure shows the requests and resulting changes side by side. Each request is recorded as a separate authoring transaction, even when several are processed in one execution.

\subsection{Routing and Extending Skills}

Reactant implements operations through Agent Skills~\cite{anthropic_agent_2025}, with each \texttt{SKILL.md} defining activation, scope, workflow, and failure behavior. Invoked with a target file, the \texttt{execute} router reads the entire document and runs the skill for the first actionable annotation in source order. After a successful operation, it submits the structured edit witness and resulting states to the kernel for commit and rereads the committed file before selecting the next annotation. It stops on failure or an unchanged, unresolved annotation.

Authors can extend the vocabulary by placing a skill under \texttt{.reactant/skills/}. The router checks this location, configured skill roots, and then the installed skills; no registry change is required. A custom skill follows the same rules for activation, scope, tag consumption, and failure detection. 

\subsection{The Transaction Kernel}

\begin{figure}[t]
\centering
\begin{minipage}{\linewidth}\sffamily
\begin{tcolorbox}[casefinal,title={1  Author's document and editor}]
\small Draft text + typed inline request + optional scope markers\hfill\textbf{Read in context}
\end{tcolorbox}
\vspace{2pt}
\begin{tcolorbox}[casefinal,title={2  General-purpose agent}]
\small Select the first actionable request\quad$\longrightarrow$\quad Load its skill\quad$\longrightarrow$\quad Edit staged files\par\smallskip
\textcolor{caseTeal}{Semantic responsibility: interpret the request, context, and scope.}\par
Submit a structured edit witness; re-read the document before the next request.
\end{tcolorbox}
\vspace{2pt}
\begin{tcolorbox}[casefinal,title={3  Shared transaction kernel}]
\small
\begin{tabularx}{\linewidth}{@{}XXX@{}}
\textcolor{caseBlue}{\textbf{Begin}} & \textcolor{casePurple}{\textbf{Commit}} & \textcolor{caseOrange}{\textbf{Inspect}} \\
Verify the reported tag; capture prior changes; create a staged mirror.
& Discover changes; validate the witness against states; store objects and versions.
& Follow validated links; query origins; open requests and recorded states.\\
\end{tabularx}
\medskip\textcolor{caseBase}{\textbf{Durable record:} request + skill identity + context hashes + before/after states + validated witness}
\end{tcolorbox}
\end{minipage}

\caption{One annotation connects semantic execution to mechanical recording. The agent interprets the request and submits a structured edit witness after execution. The kernel verifies the reported annotation and validates the witness's edit correspondences against the snapshots. The transaction retains the witness with the version transitions to support longitudinal lineage; exact snapshots remain the source for document restoration.}
\Description{Three horizontal bands show a document with an inline request, agent routing and skill execution, and the kernel's begin, commit, and lineage stages. The agent edits staged files and submits a structured edit witness containing edit correspondences. The kernel validates the witness against the before and after states and retains it with the request, skill identity, context hashes, and exact objects.}
\label{fig:architecture}
\end{figure}

History is enabled per project by initializing a \texttt{.reactant} directory, and every router dispatch opens an authoring transaction (Figure~\ref{fig:architecture}) in the directory. The router supplies the exact annotation, its interpreted skill name and prompt, the target document, and the affected passages. 

At begin, the kernel compares the current file with its recorded head. Previously untracked content becomes a \emph{base} version, whose production history is set to the default. A difference since the last recorded state becomes an \emph{unbound interval}, represented internally by a \texttt{manual} version edge, which indicates that this change happened through non-Reactant behavior (e.g., human editing, copy-paste). The kernel then captures a project manifest and creates a staged mirror. The skill receives staged paths and edits that mirror.

At commit, the kernel compares the mirror with the initial manifest to discover all changed, created, and deleted files, and validates the submitted structured edit witness against the before/after states. A citation request can therefore produce one transaction containing independent transitions for a manuscript and its bibliography. The kernel stores file contents as immutable SHA-256-addressed objects, writes version nodes and derived lineage indexes using the validated correspondences, records the transaction and its witness, installs the resulting files, and advances each affected document's head. Each document has its own version tree, and the transaction book connects files changed by the same operation.

A transaction record contains the verbatim annotation and source location, skill name and definition hash, available AI agent metadata (e.g., model name, reasoning effort), context-file hashes, status, before/after object and version identifiers, and the validated structured edit witness. A context hash identifies the recorded input state. These choices keep the record independent of a particular host's transcript format while making its capture boundary explicit.

\begin{figure}[t]
\centering
\input{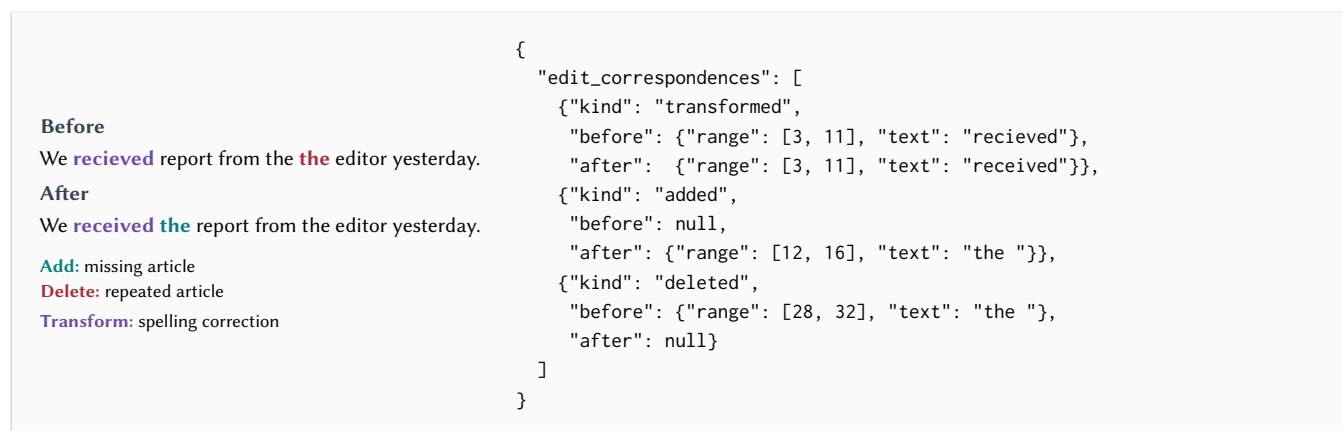}
\caption{A structured edit witness for one proofreading request. The JSON payload contains three edit correspondences: a transformation, an addition, and a deletion. The agent submits this witness with the transaction for the snapshots shown at left. Ranges are zero-based, half-open character offsets; article spans include a trailing space, and \texttt{null} denotes no predecessor or successor. The kernel validates the witness against the snapshots before retaining it.}
\Description{An untitled rectangular block shows before and after sentences at left and a JSON witness at right. Before: We recieved report from the the editor yesterday. After: We received the report from the editor yesterday. Purple marks the spelling transformation from recieved to received, teal marks the added definite article before report, and red marks the deleted second article before editor. The witness's edit correspondences record a transformed span at range 3 to 11 in both snapshots, an added span at range 12 to 16 in the after snapshot with a null predecessor, and a deleted span at range 28 to 32 in the before snapshot with a null successor.}
\label{fig:edit-correspondence}
\end{figure}

\paragraph{Structured edit witnesses.}
A structured edit witness is the payload through which the executing agent declares how it changed the document. After executing a skill, the agent submits this witness with the transaction. Its \texttt{edit\_correspondences} array contains the individual edit relations: \emph{added} spans have no predecessor, \emph{deleted} spans have no successor, and \emph{transformed} spans explicitly link earlier wording to its revision. Each correspondence identifies character ranges and their text in the before/after snapshots. Before accepting the witness, the kernel checks that the ranges and text match the actual snapshots and that the declared edits, together with unchanged text, reproduce the after-state; it then stores the validated witness with the transaction. Figure~\ref{fig:edit-correspondence} shows all three relations in one proofreading witness: adding a missing article, deleting a repeated article, and transforming \texttt{recieved} into \texttt{received}. The correction preserves a predecessor link while attributing the new spelling to the proofreading transaction. Linking predecessors across successive transactions establishes longitudinal lineage through changed wording. Here, \emph{verified} means that the witness is consistent with the recorded states; its transformation relations remain the agent's declarations, not proof of semantic correctness.

Snapshots alone underdetermine this relation: correcting a word and deleting it before independently generating a replacement can produce identical before/after text. Git provides word diffs and heuristics for following moved or copied text~\cite{git_diff_2026,git_blame_2026}, but these do not capture an executor's declared transformation links. Git could store our transaction records, yet using Git alone would still require the protocol that elicits, validates, and retains these links at each request boundary. Capturing this additional evidence during execution makes transformation lineage explicit; histories without it can only infer such correspondences after the fact.

Execution is serialized to one open transaction per project. If the real project changes while a transaction is open, commit detects the divergence and fails rather than installing the staged result over it. An ordinary skill failure also leaves its unresolved annotation in the document. A no-op can be recorded as a distinct version node with the same object, after which the router stops if the annotation remains unresolved. In our current implementation, writes are ordered to leave inspectable records, but we do not provide database-style crash-atomic installation across multiple files. An integrity command reports inconsistencies rather than reconstructing a missing history.

\subsection{Inspecting History with Reactant Replay}\label{sec:replay}

Reactant Replay is a browser-based viewer for inspecting recorded document histories (Figure~\ref{fig:intro-lineage}). Its timeline provides access to snapshots, with word colors indicating recorded origins and comparisons showing revisions. A word's \emph{Origin} links it to the operation and exact request that introduced it, alongside the surrounding before/after text. Initial text and unbound changes remain explicitly labeled when no recorded request explains their origin.

The \emph{Trace} view follows a selected word or passage across versions, connecting earlier formulations to the requests that changed them. It summarizes unchanged intervals and links each step to its snapshot or before/after comparison. These views support inspecting both the origin of particular wording and the development of a passage, as demonstrated in Section~\ref{sec:cases}.

\section{Tracing This Paper's Introduction}\label{sec:cases}

When several requests shape the same paragraph, what can an author recover from its history? Here we demonstrate a passage edited by Reactant and use Reactant Replay to show how an author can inspect a word's origin in a chosen snapshot and trace a passage across its recorded revisions. The example comes from the recorded history of this paper's own introduction (Yes, we are using Reactant to help draft its own paper!).

The paragraph beginning ``Beyond provenance and lineage'' introduces three practical advantages of Reactant. Its editor-independence passage now connects the ability to retain a preferred editor to findings from prior research. The final wording presents this connection as one continuous argument. Its history shows how the argument, the source, and the role of that source developed through separate requests.

\begin{figure}[tp]
\centering
\includegraphics[width=\linewidth]{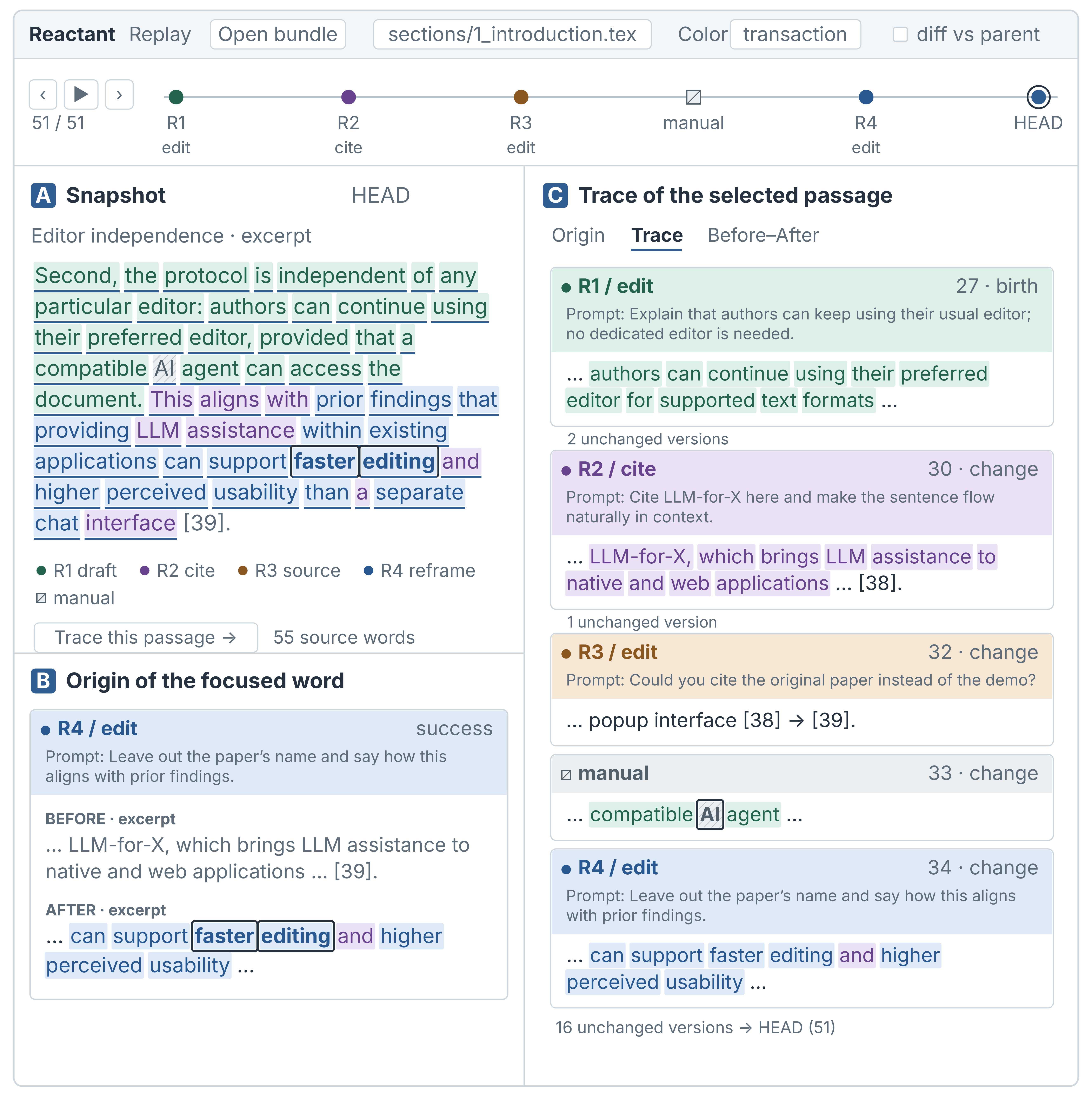}
\caption{Reactant Replay inspecting this paper's introduction. The timeline selects a snapshot, whose word colors identify recorded origins. The Origin view connects ``faster editing'' to R4's request and its before/after wording. Tracing the editor-independence passage shows its development from R1's drafting through R2's citation insertion, R3's source correction, a manual change, and R4's reframing, followed by unchanged versions.}
\Description{A Replay interface has a transaction timeline above two columns: the document snapshot and Origin details on the left, and a chronological Trace on the right. The snapshot highlights faster editing. Its Origin shows R4 replacing a system description with a research finding. Trace of the selected passage begins with R1's inline drafting request, then shows R2 adding a cited example, R3 changing to the original publication, an addition of AI labeled manual, and R4 reframing the source as evidence. Each operation card contains its prompt in smaller gray text. Later unchanged versions are summarized.}
\label{fig:intro-lineage}
\end{figure}

We selected this passage because its revisions bring together drafting, citation insertion, source correction, and reframing. The case ends at a snapshot saved before we drafted this section. Its full ancestry contains 62 recorded states; the interval from the first selected request's result to this endpoint contains 25 states, comprising 14 transaction-backed states and 11 unbound states. Each request also preserves its input state. Figure~\ref{fig:intro-lineage} shows the passage at this endpoint, retaining R1--R4 as chronological labels for four focal requests.

Across the introduction's full recorded ancestry up to this endpoint, 38 successful skill invocations were recorded: 18 \texttt{edit}, 8 \texttt{proofread}, 6 \texttt{ph}, 4 \texttt{cite}, and 2 \texttt{plan}. Twenty of these invocations used Codex (with GPT-5.6-sol) and 18 used Claude Code (with Opus 5). The remaining states comprise one initial \emph{base} state and 23 \emph{unbound} states recorded between invocations. These counts include revisions outside the selected passage.

The timeline lets the author move between snapshots. In the displayed snapshot, a word's Origin shows its originating operation, original prompt, and before/after revision. Selecting the editor-independence passage and choosing \emph{Trace this passage} brings its history into a chronological sequence, including earlier formulations that have since been replaced. Each trace step opens the corresponding snapshot for inspection, while unchanged intervals are summarized between changes.

\paragraph{Starting with the current claim.}
Suppose an author clicks ``faster'' in the phrase ``faster editing'' to examine how this claim entered the paragraph. Its Origin opens R4, an \texttt{edit} request: ``Leave out the paper's name and say how this aligns with prior findings'' The after-state connects editor independence to findings about editing speed and perceived usability. The before-state instead describes LLM-for-X and its popup interface. The current claim thus opens onto an earlier formulation whose system name and description have disappeared from the working passage.

\paragraph{Tracing the passage's development.}
Selecting the whole editor-independence passage opens a longer history in Trace. Its first step is R1's \texttt{edit} request for three advantages: staying within the writing document, retaining a preferred editor, and extending the operation vocabulary. At this point, an inline instruction occupies the position of the future paragraph. Execution turns it into prose, including the initial editor-independence sentence.

R2 shows where the description and citation entered the passage. Its \texttt{cite} prompt asks to add LLM-for-X in context and smooth the sentence. Before execution, the passage contains the editor-independence argument without this supporting example. After execution, it describes the system's integration into native and web applications and cites the demonstration publication~\cite{teufelberger_demonstrating_2024}. The transaction connects the prose addition to the accompanying bibliography change.

The next focal request, R3, asks to cite the original paper instead of the demo paper. Its before/after comparison preserves the same descriptive sentence while changing the citation key to the original LLM-for-X paper~\cite{teufelberger_llmforx_2024}, with a corresponding bibliography update. The trace separates this source correction from R4's later reframing of what the publication should support. In general, these steps recover the changing relationship between argument and evidence: an initial rationale acquires a cited example, a corrected source, and a different role for that source in the finished argument. This trace lets either the author themself or a coauthor later recover the requests and revisions that shaped a finished claim, addressing the problem of losing the intent behind successive edits when only the final wording remains, as well as distinguishing which parts are extensively modified by AI, which are only proofread, and which are from external editing (e.g., writing by humans or copying from other sources).

\section{Reactant in Practice}\label{sec:evaluation}
Since Reactant was developed, our lab has used it to assist with preparing four research paper submissions, course materials for two different classes, and one grant proposal. We have also invited a few colleagues to try it out. Three of them have used it for over four months; all three are academic researchers with more than six years of experience writing papers and have agreed to share their experiences with and feedback on our system.
Throughout this period, we maintained informal contact, primarily through face-to-face conversations, to discuss their experiences and gather feedback. Their experiences connect the AI-assited writing practices to recovering earlier decisions, continuing work across different AI agents, and developing new uses of the transaction history. Each of the three researchers reported using Reactant for at least 100 transactions across at least three different projects during this period. 
We report these accounts as situated examples of use rather than as a summative evaluation of Reactant's usability or effectiveness.

\subsection{Recovering the Request behind a Revision}\label{sec:case-e1}

All three users mentioned that inline annotations fit their habit of leaving comments in \LaTeX{}: some comments are notes to self, while others ask coauthors to fix a passage. Reactant extends this familiar practice to requests for an agent. E1 particularly valued being able to place these requests within the writing document without switching windows. E1 also developed a VS Code feature that automatically runs \texttt{/execute} in the AI agent when a keyboard shortcut is pressed, further reducing the effort of switching between windows. We later incorporated this feature into Reactant.

The value of retaining these requests became concrete when E1 used Replay to inspect a passage in a paper coauthored with a student (who was also using Reactant). According to E1, they examined the recorded request alongside the revisions to understand the student's stated intent and what AI agent had changed. E1 traced the puzzling passage to an earlier AI instruction that had been too vague and had led the agent to change content in another file as well. The records helped E1 understand how the passage had developed during the student's writing process.

E1 also reported that reviewing the history made it possible to distinguish their own writing, text that AI had proofread, and passages generated entirely by AI. This account highlights the importance of preserving the kind of intervention made in a passage. Proofreading an author's wording and generating a passage involve different forms of delegation, even though both involve AI. For E1, the history made these differences visible when returning to the finished text.

\subsection{Continuing Work across Agents and Tasks}\label{sec:case-e2}

E2 reported frequently switching between agent platforms due to rate limits or differences in model capabilities. E2 valued that Reactant's model-agnostic history persisted across these switches. 

E2 also valued the version-management features, particularly the ease of comparing the document before and after a prompt and comparing non-adjacent versions. E2 described revising a passage through several prompts and multiple skills, then comparing the resulting wording with an earlier version. This comparison showed the cumulative effect of the intervening operations on the passage. 

E2 also valued customization and reported frequently defining skills for particular tasks. The ability to define an operation's behavior let E2 adapt Reactant's vocabulary to the work at hand, alongside general operations such as proofreading. For example, during course preparation, E2 defined a conversion skill to turn concepts from technical papers into explanations that the course audience could understand. 

\subsection{Building New Uses from the History}\label{sec:case-e3}

E3 described a creative appropriation of Reactant, adapting it for many uses and creating new skills. One example began with a request to an AI agent to analyze E3's transaction history, identify frequently repeated prompts, and consider whether they could be extracted into skills. The agent identified recurring instructions to avoid particular words and em dashes, proposed a dedicated writing-style skill, and generated it. The resulting skill captured preferences that E3 had repeatedly expressed during earlier revisions, including avoiding words E3 rarely uses in their own writing. E3 found this experimentation enjoyable and described the transaction records themselves as a substrate for further functionality. The style skill was one instance of E3's broader practice of extending Reactant through custom skills and new ways of using its records.

E3 also described a recurring sequence for investigating a sentence's history. Most of the time, E3 first asked an AI agent to inspect the transaction records and explain why the sentence had been written that way, including the prompt used at the time. E3 then opened Replay to confirm the answer against the recorded history. E3 found this useful for recovering the context of earlier writing decisions. 

\section{Discussion and Conclusion}\label{sec:discussion}

Reactant brings agentic assistance into authors' existing writing practices while preserving a history that remains useful beyond the immediate revision. Inline requests keep instructions beside the relevant text, and the verified transaction protocol allows authors to change agents or introduce custom skills while continuing the same history. Recording each request with its exact snapshots and validated structured edit witness makes both the delegated operation and the relationships declared in its edit correspondences available for later inspection. These properties make Reactant an extensible substrate for agentic authoring: authors can revisit decisions through longitudinal lineage, collaborators can inspect how a passage developed, and agents can use earlier requests to support subsequent work.

Our use cases and colleagues' experiences are concentrated in academic writing by experienced researchers. They illustrate how Reactant can fit ongoing work, but we have not conducted a formal comparative usability study. We therefore cannot establish whether Reactant is easier to learn, more efficient, or preferable to conventional chat-based or editor-integrated workflows. The contribution demonstrated here is the capability to retain individual request boundaries and explicit transformation links across compatible agents and custom operations, which is not captured from a conventional chat transcript or file-version history alone.

In the future, we plan to formalize the features emerging from E3's use: extracting reusable skills from recurring instructions and supporting conversational inquiry into a document's history. A skill-authoring interface could present candidate preferences alongside the transactions from which they were inferred, allowing authors to revise them before adopting a skill. A history assistant could connect each explanation directly to the relevant requests and versions in Replay, making it easy to inspect its basis. Building on these features, we will study how writers develop and share custom skills, including the potential for a community skill library, and evaluate how these tools affect authoring and retrospective review. More broadly, we will explore applications beyond manuscript authoring. For example, annotations in a visual design brief could direct successive layout or illustration revisions, allowing designers to compare alternatives and revisit the instructions behind them. These scenarios would let us investigate how the transaction substrate supports the development of interpretations and design decisions across a project. We hope that transactional inline annotations can become a general paradigm for human--AI interaction, enabling people to direct AI within their ongoing work and to inspect, question, and build on the changes it makes.

\bibliographystyle{ACM-Reference-Format}
\bibliography{main}

\end{document}